# Interplay Between Mixed and Pure Exciton States Controls Singlet Fission in Rubrene Single Crystals

**Dmitry R. Maslennikov[1], Marios Maimaris[1], Haoqing Ning[1], Xijia Zheng[1], Navendu Mondal[1], Vladimir V. Bruevich[2], Saied Md Pratik[4], Andrew J. Musser[3], Vitaly Podzorov[2], Jean-Luc Bredas[4], Veaceslav Coropceanu[4], Artem A. Bakulin[1]\***

[1] Department of Chemistry and Centre for Processible Electronics, Imperial College London, UK
[2] Department of Physics and Astronomy, Rutgers University, New Jersey 08854, USA
[3] Cornell University, Department of Chemistry and Chemical Biology, USA
[4] Department of Chemistry and Biochemistry, The University of Arizona, Tucson, Arizona 85721-0041, USA
\* a.bakulin@imperial.ac.uk

**Abstract:** Singlet fission (SF) is a multielectron process in which one singlet exciton **S** converts into a pair of triplet excitons **T+T**. SF is widely studied as it may help overcome the Shockley-Queisser efficiency limit for semiconductor photovoltaic cells. To elucidate and control the SF mechanism, great attention has been given to the identification of intermediate states in SF materials, which often appear elusive due to the complexity and fast timescales of the SF process. Here, we apply 10fs-1ms transient absorption techniques to high-purity rubrene single crystals to disentangle the intrinsic fission dynamics from the effects of defects and grain boundaries and to identify reliably the fission intermediates. We show that above-gap excitation directly generates a hybrid vibronically assisted mixture of singlet state and triplet-pair multiexciton **[S:TT]**, which rapidly (<100fs) and coherently branches into pure singlet or triplet excitations. The relaxation of **[S:TT]** to **S** is followed by a relatively slow and temperature-activated (48 meV activation energy) incoherent fission process. The SF competing pathways and intermediates revealed here unify the observations and models presented in previous studies of SF in rubrene and propose alternative strategies for the development of SF-enhanced photovoltaic materials.

## Introduction

Coherent dynamics are identified to underpin ever more ultrafast processes in molecular photochemistry, from biological light harvesting[1–3] to donor-acceptor charge transfer[4,5] to molecular exciton multiplication processes.[6–9] Frequently these processes entail conversion between optically bright and dark electronic states and the coherence is derived from strong coupling to vibrational modes. Such vibronic coupling induces mixing between bright and dark states, with these mixed states serving as key transition states or intermediates along the potential energy landscape. This framework has been widely applied to singlet fission (SF), in which a photoexcited singlet state **S** converts into a pair of triplet states **T+T**, mediated by a spin-entangled multiexciton state **TT**.[8–18] The latter state exhibits key hallmarks of its mixed character—combinations of singlet-, charge-transfer-, and triplet-derived spectral signatures;[19–21] stabilization relative to the uncoupled parent states;[11] and ability to emit photons and diffuse over micron length scales despite being nominally dark and localized.[22–25] Whether exothermic or endothermic, its formation on ultrafast timescales is found to be driven by coherent vibronic coupling.[8,9,17,26–28]

A corollary of this mechanism is that this multiexciton character should likewise be mixed into the photoexcited **S** state, though not necessarily at the Franck-Condon point. Direct mixing of **TT** into **S** is predicted theoretically in several systems,[9,27,29,30] but this effect has only been inferred experimentally from the observation of coherent dynamics.[9] In the best known example, the direct excitation of a coherently mixed **[S:TT]** state was initially proposed in pentacene and tetracene based on the presence of low-energy signatures in transient photoelectron spectroscopy.[6,7] However, subsequent analysis[31] and recent angle-resolved photoelectron spectroscopy[32] indicate these systems follow the standard conversion from a pure **S** state into mixed **TT**. It thus remains unclear whether the **[S:TT]** state can be directly photoexcited, despite its established ability to emit photons.[11,19,22,23,33]

We address this question using single crystalline rubrene of very high purity. Rubrene boasts nearly isoenergetic **S** and **TT** states and long triplet diffusion lengths,[34,35] making it optimal for SF applications. However, the $C_{2h}$ symmetric stacking in rubrene results in negligible electronic coupling between **S** and **TT**,[35] resulting in especially pronounced sensitivity to static and dynamic disorder. The dynamics of SF in rubrene have thus been controversial, with reported mechanisms ranging from <100 fs coherent to >10 ps incoherent channels.[14,36–44] These early studies were limited by the complex interplay between intrinsic fission dynamics related to the crystal structure and extrinsic dynamics tied to defects and grain boundaries. In addition, in the leading time-resolved studies, the assignments to **S** and **TT** states were limited by the absence of distinct spectral fingerprints not overlapped with the ground-state absorption or thermal artefacts.[14,36,45]

Here, we study high-purity single crystals using sub-10-fs-to-ms transient absorption spectroscopy in the near-infrared, a background-free region where unique fingerprints of **S**, **TT**, and a mixed **[S:TT]** state can be readily distinguished. We recover dynamics of slow, thermally activated SF from **S** similar to prior reports.[37–42] However, our key finding is that **S** is not the initial photo-excited state. Instead, the initial species contains unique signatures of a mixed **[S:TT]** state. Thermalization causes this state to collapse either into **TT**, which leads to a fast fission channel, or into **S**, with the latter population responsible for the slow fission channel. The branching from **S:TT** and its dependence on the excitation frequency reconciles the competing claims of coherent and incoherent fission in rubrene and demonstrates that even dark, multiexciton states can be directly photoexcited through strong vibronic coupling.

## Results

Rubrene single crystals were grown using the physical vapor transport (PVT) technique, as it is well-established to provide molecular crystals with a very low concentration of defects, thereby minimizing the influence of static disorder on charge and exciton dynamics.[23,46] Previous work on tetracene has demonstrated that the fission rate and mechanism inferred from polycrystalline films are dominated by grain boundaries and other types of disorder and differ significantly from those in single crystals.[47,48] To distinguish the intrinsic SF properties from extrinsic effects, it is therefore imperative to study low-defect crystalline materials. We note that the photoluminescence and other photophysical properties of organic solids, even in their highly purified single crystalline form, are very sensitive to trace impurities and structural defects,[49–53] thus requiring special care during the material growth. Thus, for this study, we have prepared exceptionally pure PVT-grown rubrene single crystals similar to those used in the recent first demonstration of a photo-Hall effect in organic semiconductors.[54]

Figure 1a shows a photograph of a representative PVT-grown rubrene single crystal studied in this work, with the corresponding molecular packing motif schematically overlaid on the image.

The crystal is a hexagonally shaped thin plate, with its largest facet corresponding to the bc crystallographic plane (the so-called high-mobility plane of orthorhombic rubrene). The hexagonal habitus was used to match the crystal axis orientation with the shape of the crystal (Section I of SI). Figure 1c shows the absorption spectra of the crystal for the incident light linearly polarized along the 0C and 0B crystallographic axes.

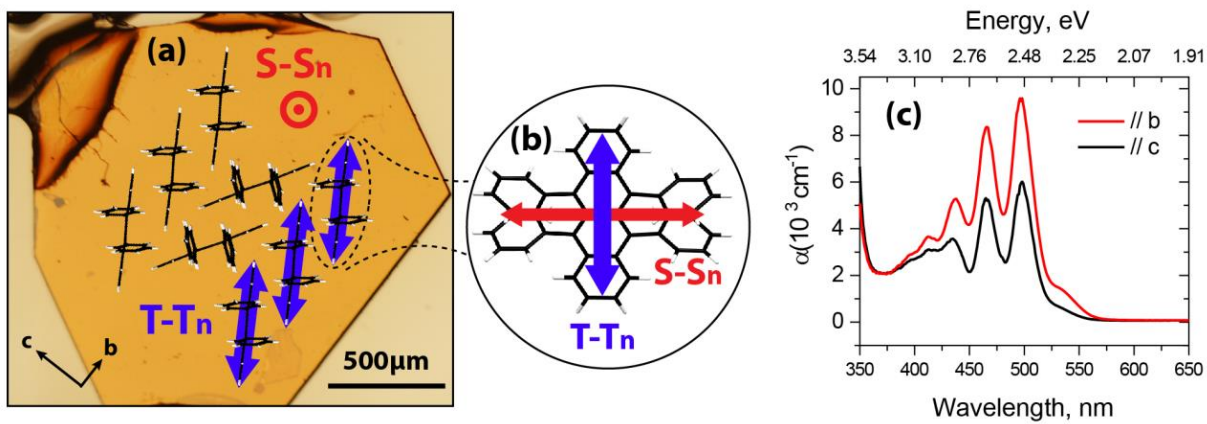

*Figure 1.* a) *Microscopy image of the bc-plane upper facet of a rubrene single crystal in correspondence with the molecular-packing schematics. b) Illustration of a rubrene molecule with the blue [red] arrow indicating the T-T$_n$ [S-S$_n$] transition dipole moment direction. c) UV-VIS absorption spectra of a rubrene single crystal for the photoexcitation polarized along b and c axes of the crystal.*

We addressed exciton dynamics in the crystal using transient absorption spectroscopy (TAS) on three instruments with temporal resolutions of <10 fs, 200 fs, and 10 ns. The results were obtained using probes in the near-infrared spectral range, which is uncontaminated by ground-state absorption and thermal artefacts.[55,56] Moreover, this range contains unique fingerprints of the key states in the SF pathway in many other acenes: narrow vibronic resonances for **T-T$_n$** transitions and broad bands for **S-S$_n$** transitions.[26,57–59] In crystalline rubrene, these states can be further distinguished due to the significant anisotropy that arises from the herringbone molecular packing[46,60], and the orthogonal transition dipole moments of the **S-S$_n$** and **T-T$_n$** transitions (Fig 1b). To balance the signal coming from the photoinduced absorption (PIA) of singlet and triplet excitons, the probe polarization was oriented approximately at 60° relative to the 0B crystal axis (± 5° due to manual positioning of the crystal).

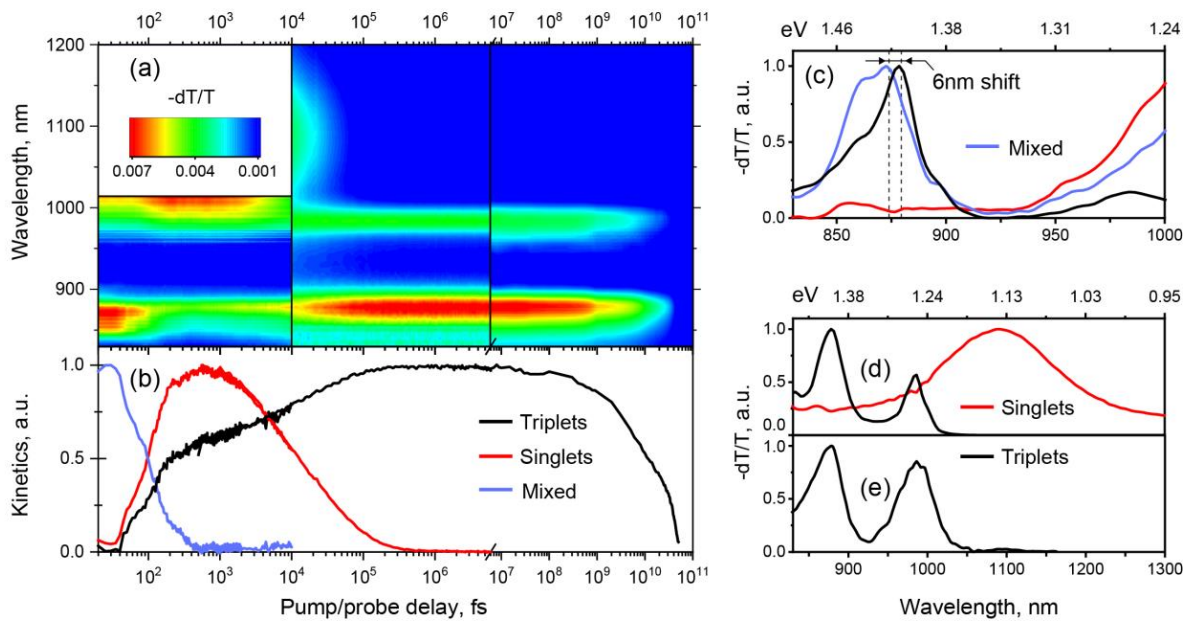

**Figure 2.** *Room temperature transient absorption in a rubrene single crystal measured using three different setups (a) with temporal resolutions of <10 fs, 200 fs, and 10 ns. Kinetics (b) and spectra (c-e) of the **S**, **T+T**, and **[S:TT]** states provided by the global analysis of the experimental datasets.*

Figure 2a presents full TAS data from the single crystal shown in Figure 1a. The results from the three setups were scaled and merged to capture the full excited-state evolution from 10 fs to 50 µs. The probe region of sub-10-fs experiments was narrower than for other experiments due to the technical limitations of supercontinuum generation. Slight shifts of shared spectral features indicate this procedure is accurate to within 5 nm. On the ns/µs timescale we detect a single species with 13 µs lifetime, characterised by two sharp transitions at 878 nm and 985nm separated by a 1300 cm⁻¹ spacing. These features are widely observed in solutions and thin films of polyacene derivatives[20,21,26,57,61,62] and the lowest-energy peak closely matches our DFT-calculated gap between relaxed **T** and **Tₙ** states. We thus assign these to a vibronic progression of **T-Tₙ** transitions of either free triplets **T** or electronically decoupled triplet pairs **T+T**. We directly observe the formation of this state within 100 fs and the slow (>ps) part of the signal growth matches the decay of a broad photoinduced absorption beyond 1000 nm. The latter band is typical of the **S** state in numerous polyacenes[57–59,61,62] and reveals slow SF, consistent with the incoherent SF model.

However, we find that **S** is not the initial photoexcited state. On the <100-fs timescale, we resolve a distinct species with a sharp band centred at 865 nm. Closer examination reveals this feature is associated with a broad photoinduced absorption >1000 nm, and its decay tracks the rise of **S**. This qualitative analysis indicates a two-step progression: From an unknown state to **S** and from **S** to **T+T**. To identify the initial state and disentangle its interplay with the other species, we spectrally decompose the data in each temporal range using a combination of singular value decomposition and a genetic algorithm. This method permits extraction of the characteristic spectral species without any pre-determined kinetics scheme.[63] The resulting population kinetics and spectra are presented in Figure 2b. We find good agreement between the extracted **T+T** spectra in all ranges and the **S** spectra corroborate the analysis above. The initial species contains a mixture of characteristics of the **T-T$_n$** and **S-S$_n$** transitions. However, it cannot be described as a simple linear combination of singlet and triplet states; for instance, the **T**-like band is shifted by 6 nm relative to **T+T**. Based on the hybrid character of this spectrum, we assign the initial species to a mixed **[S:TT]** multiexciton state. Similar hybrid states have been identified as the immediate product state of SF in solutions and thin films[16,19,22,26,61] and some works identify the blue-shift of **T-T$_n$** relative to **T+T** as a signature of the multiexciton binding energy.[11,19] However, this is the first explicit observation that this dark mixed state can be directly photoexcited.

The population dynamics reveal two channels of triplet formation. The photoexcited **[S:TT]** state decays with ~ 80 fs time constant, resulting in parallel formation of independent **S** and **TT** components. This timescale is typical of vibrational relaxation and is consistent with the above-gap, vibrationally hot excitation in our <10-fs experiment. Such branching should follow a coherent mechanism, with relaxation along the mixed **[S:TT]** potential energy surface into **S** or **TT** configurations weighted by their contribution to the initial wavefunction.[64] Under 400-nm excitation conditions, this coherent channel provides ~ 40% of the final triplet yield. Subsequently, the **S** population peaks at ~200 fs and undergoes further SF with a time constant of ~10 ps. This slow fission rate suggests that **S** requires energy to overcome a potential energy barrier to convert back to the mixed state **[S:TT]** or directly into **TT**. That is, the relaxed **S** is stabilized relative to **TT** and does not couple to it. To evaluate the energetics of the slow SF channel, we performed further temperature-dependent measurements in this regime.

Figure 3 shows the temperature dependence of the extracted **S** and **T** dynamics and spectra. The kinetics (Figure 3a, b) demonstrate that fission slows dramatically at low temperatures, to a time constant of 150 ps at 125 K. This result confirms the endothermic nature of the picosecond channel of SF in rubrene; though **S** and **TT** are nearly isoenergetic,[14,35,36] *relaxed* **S** faces a sufficient barrier to **[S:TT]** or **TT** to significantly impact the dynamics. Upon cooling, we further observe that the spectral signatures of *S* and *T* narrow at a similar pace (Figure 3c,d and Section II of SI). At the same time, the *S* peak undergoes a blueshift, which is not observed in the *T* peaks. The energy spacing between the *T* peaks remains constant at 1300 cm⁻¹ and matches the energy of the vibrational mode most coupled to electronic states, observed in Raman measurements (Figure 3e). Following previous studies, we attribute the narrow triplet signatures to significantly different degrees of delocalization for singlet and triplet excited states in organic single crystals.[65,66]

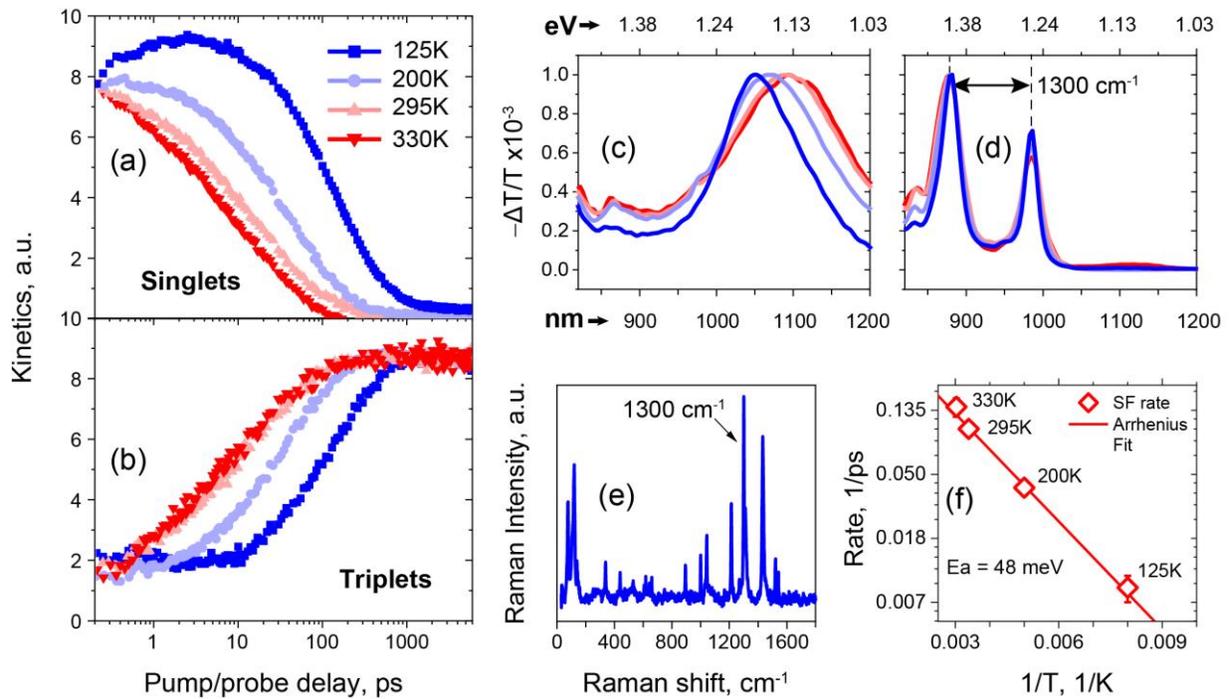

***Figure 3***. *Kinetics and spectra of singlets (a, c) and triplets (b, d) from the global analysis of the temperature-dependent transient absorption spectroscopy of rubrene single crystals (430nm excitation). Raman spectra of a rubrene single crystal measured with the excitation polarization directed along the 0B axis (e). SF rates in rubrene single crystals as a function of inverse temperature (f).*

The magnitude of the barrier between relaxed **S** and **[S:TT]** (activation energy $E_a$) can be estimated by applying the Arrhenius model to the fission dynamics. As shown in Figure 3f, the logarithm of the fission rate is inversely proportional to temperature over the entire measured range, indicating the thermally activated model is appropriate. Our fit yields an activation energy of 48 ± 3 meV, which is in reasonable agreement with values previously reported for more disordered rubrene systems.[14,40,42,43,67]

Figure 4a presents a state model of SF in defect-free single-crystalline rubrene based on our TAS findings, which reconciles the existing literature data. Excitation with above-optical-gap photons leads to direct formation of mixed **[S:TT]** states. **[S:TT]** are partially bright states as they borrow transition dipole moment from bright S.[22] Having both S and T character, the **[S:TT]** states allow transitions into the **$T_n$** and **$S_n$** manifolds and thus possess both **S**-like and **T**-like spectral signatures in their transient absorption.[21] The **T-$T_n$** transition is, however, red-shifted in **[S:TT]** compared to **T** or **T+T** due to the multiexciton stabilization energy.[11] In pentacene derivatives, **[S:TT]** mixing was previously shown to be promoted by the overlap of **S** and **TT** vibronic manifolds and we believe this mechanism is also present in tetracene derivatives including rubrene.[9]

Within the first ~100fs after photoexcitation, the initially formed **[S:TT]** mixed states relax to either predominantly triplet-pair **TT** (and then to **T+T**), or to 'pure' singlet states (we estimate ~2.7 times more relaxation into **S** compared to **T,** based on the relative amounts of prompt SF). This exothermic process is the first, fast, and most likely coherent SF pathway in rubrene. Cooling to the pure **S** is followed by slow (>10ps) thermal activation of singlets over the 48 meV barrier back to **[S:TT]** and further to triplet states. Our data show that the rate of this second process is limited by thermal activation and we do not detect any further intermediates prior to **T+T**. Therefore, the conversion of **TT** to **T+T** likely occurs on a faster timescale than 10ps.

Crucially, the presented model suggests a simple way to control the branching between coherent and incoherent SF pathways. The initially excited states must be bright and therefore have substantial **S** character. However, **S** and **TT** states exist in a dense vibronic manifold and one can expect that the higher-lying states will have higher degree of mixing and higher contribution of **TT** character. This higher triplet character should increase the probability for the mixed state to branch off directly to pure **TT** via the coherent fission pathway. Indeed, we find that excitation with higher photon energy increases the relative contribution of the coherent pathway and decreases that of activated fission via **S** (Figure 4b,d). While the general dynamics of the two SF pathways stay the same, importantly, the contribution to the total triplet yield from the ultrafast coherent process changes from ~17% to ~40% over our pump energy scan.

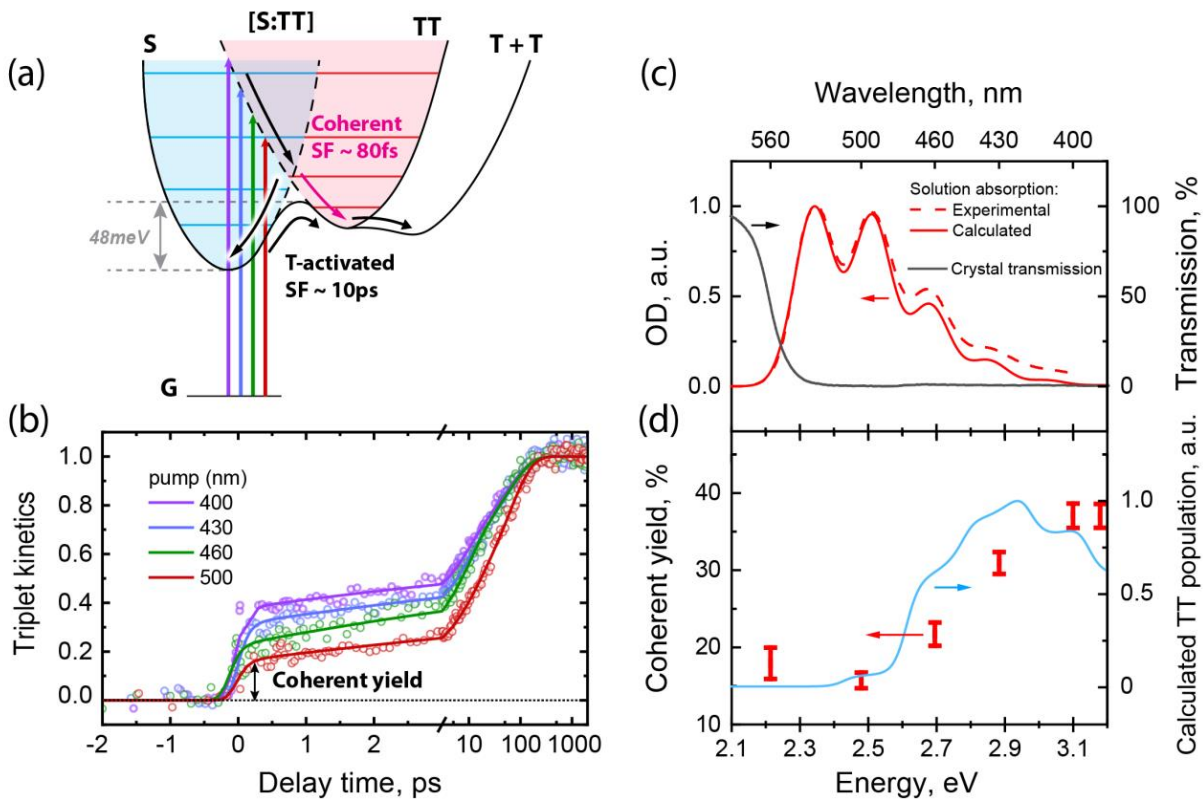

***Figure 4.*** *a) SF model for rubrene single crystals. Tuning of the excitation wavelength allows ultrafast control of the triplet states generated immediately after the excitation (coherent SF). Thick blue lines indicate photoexcitation with different pump wavelengths; curved black and pink lines point to internal interband transitions. b) Triplet SF kinetics dependence on the pump wavelength. c) Absorption spectrum of rubrene molecules in deuterated chloroform solution as calculated with the 3-state model and comparison with experiment. Experimental transmission spectrum of the crystal is shown for reference. d) Relative increase in **TT** state population as a function of excitation pulse energy compared with the calculated coherent triplet yield.*

The analysis given above is further supported by the results of electronic-structure calculations and vibronic modelling. We first estimated the reorganization energy related to the **S** →**TT** transition. This energy can be roughly approximated as the sum of the relaxation energies associated with the **G→T** and **S→T** electronic transitions, where **G** denotes the ground state. The calculations performed at the density functional theory (DFT) level with the global hybrid B3LYP functional and 6-31G(d,p) basis set yield a reorganization energy of about 0.3 eV. However, our previous studies[68] have shown that B3LYP-based calculations usually underestimate the relaxation energies by up to 30%. In addition, we neglected the contribution to the reorganization energy due to the interactions with surrounding molecules; for non-polar systems, this contribution is typically in the range of 0.1−0.2 eV.[69] Taking all these factors into account, the reorganization energy related to the **S→TT** transition is estimated to be in the range of 0.4−0.6 eV. If we were to follow the classical Marcus theory, such values would lead to an activation barrier between the relaxed *S* and *TT* states on the order of 100−150 meV, *i.e.,* two-to-three times larger than the experimental 48 meV value. However, the DFT calculations indicate that the bulk of the reorganization energy is due to non-classical (high-frequency) vibrational modes, which calls for the application of the expanded Marcus-Levich-Jortner (MLJ) model. Using this model, taking into account the DFT-derived vibrational couplings, and assuming a contribution of 0.1 eV to the classical reorganization energy due to the surroundings, we estimate (Section III of SI) the activation energy to be 52 meV, in very good agreement with experiment. In order to reproduce the experimental rate constants, our calculations have to assume a relatively small value of 7 meV for the electronic coupling between the **S** and **TT** states.

To describe the hybridization of high-energy vibrational modes of the **S** and **TT** states, we consider a fully quantum-mechanical vibronic model[70] based on three electronic states (S, TT, and G) and two effective vibrational modes. This model treats the electronic and electron-vibrational interactions exactly,[70] Section III of SI. Figure 4c,d presents the results of the vibronic simulations for the rubrene absorption spectrum and the contribution of the triplet **TT** state in the **[S:TT]** mixed state populated by photon absorption as a function of photon energy. Our results confirm that the direct (coherent) generation of the **TT** state in rubrene (where the reorganization energy is large) increases upon photo-exciting the vibronic and hybridised sub-levels of **S** at larger photon energies. Our calculations capture well the trend observed experimentally (Figure 4d).

We expect that a better agreement with the observed data could be achieved if additional electronic states are included into the model. We note, for instance, that the ionization potential and electron affinity energies of solid state rubrene are 5.4 eV and 2.7 eV, respectively.[71,72] These values suggest that, when considering the electrostatic interactions between hole and electron, the lowest charge transfer (CT) state in rubrene will appear at an energy lower than 2.7 eV. As a result, the vibrational levels of the CT state could mix with the excited vibrational manifold of the **S** and **TT** states and contribute to **TT** generation at high photo-excitation energies.

Our work potentially resolves the controversy between the two SF models for crystalline rubrene that have been debated in the literature. Coherent SF is occurring immediately after photoexcitation, as highlighted for example by Miyata *et al*., and the incoherent (hopping-like) SF is taking place over a potential energy barrier via thermal activation.[14] According to the coherent SF model, an electronic wave packet formed by photoexcitation may evolve into either cold singlet or triplet-pair states via a conical intersection of the **S** and **TT** potential energy surfaces enabled by a symmetry-breaking mode; this was corroborated by the experimental observation of temperature-independent step-like PIA at 510 nm and 800 nm in the TAS kinetics, attributed to a fast (sub-100fs) formation of triplets. Our results suggest that the step-like signals may be coming from the mixed states, which possess both **S** and **T** TAS signatures. The use of high-quality samples, high temporal resolution, and wide probe range allowed differentiating **TT** and **[S:TT]** states and identifying individual states in the dynamics. We note that another explanation of the rapidly appearing triplet signal has been proposed by Turner *et al.* in the framework of incoherent SF;[36] according to that model, the rapid observation of triplet pairs with 2D electronic spectroscopy is related to weakly coupled but nearly resonant electronic energy levels of the singlet and triplet-pair states. However, this proposition, despite its attractive simplicity, is unable to explain the temperature-dependent fission rates. The scenario we put forward resolves the contradictions among the various theories by showing that both ultrafast exothermic and slow endothermic SF pathways stem from the **[S:TT]** state.

To conclude, with muti-timescale TA spectroscopy we have tracked down the SF dynamics in high-purity rubrene single crystals, from photoexcitation (~10 fs) to triplet exciton decay (~μs). Our experiments revealed that SF in rubrene begins with the formation of a very short-lived state with a hybrid singlet/triplet character, which rapidly (~80fs) coherently decays into either a singlet state or a **T+T** triplet pair. This hybrid state can be described as a vibronically-enhanced mixture of the **S** and **TT** multiexciton states. Beyond reconciling the disparate models of singlet fission in crystalline rubrene, our observation of direct excitation of a mixed **[S:TT]** state points to new ways in which SF can be optimized in endothermic systems. Here, we have demonstrated the SF pathway control via the choice of excitation photon energy. Further work may target different molecular design aspects, such as intermolecular couplings and the density of vibronic manifold, to achieve a better control of the fission mechanism and the realisation of SF-enhanced photovoltaic materials.

### Acknowledgements

We thank Hikmet Najafov for providing the absorption spectra of rubrene single crystals and Dmitry Igorevich Dominsky for helping us assigning crystallographic axes orientation with the shape of the crystal. AAB acknowledges Royal Society for the support via University Research Fellowship. DM was supported by the Imperial College London President's PhD Scholarships. The work at the University of Arizona was funded by the Office of Naval Research, Award No. N00014-20-1-2110. VP and VB thank the financial support of their part of work at Rutgers University through the Donald H. Jacobs Chair in Applied Physics.

# Interplay Between Mixed and Pure Exciton States Controls Singlet Fission in Rubrene Single Crystals

## Supporting Information

**Dmitry R. Maslennikov[1], Marios Maimaris[1], Haoqing Ning[1], Xijia Zheng[1], Navendu Mondal[1], Vladimir V. Bruevich[2], Saied Md Pratik[4], Andrew J. Musser[3], Vitaly Podzorov[2], Jean-Luc Bredas[4], Veaceslav Coropceanu[4], Artem A. Bakulin[1]***

*[1] Department of Chemistry and Centre for Processible Electronics, Imperial College London, UK*
*[2] Rutgers University, Physics Departments, USA*
*[3] Cornell University, Department of Chemistry and Chemical Biology, USA*
*[4] Department of Chemistry and Biochemistry, The University of Arizona, Tucson, Arizona 85721-0041, USA*

*\* a.bakulin@imperial.ac.uk*

## Methods

*Single crystal growth*

Rubrene single crystals were grown by multiple recrystallizations of a rubrene powder purchased from GFS Chemicals using the PVT method in a stream of ultra-high purity (UHP) He gas. Details of the crystal growth can be found elsewhere.[1,2]

*Transient absorption (TA)*

To achieve both high temporal resolution and broad time range, TA of rubrene single crystals was measured using three different setups with different time resolution and pump/probe delay range and then stitched together according to the same features in the overlapping time delay regions. The sub-10fs resolution setup used for measurements (<10ps) is based on the chirped mirrors compressor system described in ref. [3]. A commercially available TA spectrometer Helios (Spectra Physics, Newport Corp.) was used to measure the TA in the intermediate time delay region (>0.2ps, <6ns). Finally, a home-built TA spectrometer in conjunction with a q-switch nanosecond laser (Picolo, InnoLas) was used for TA measurements >1ns. For Figure 2 in the main text, we used 400nm (50nJ), 430nm (160nJ) and 532nm (75nJ) pump excitation correspondingly focused on the samples with a beam size of around 0.5 mm$^2$. As a reference for the wavelength scale we used the intermediate time delay dataset (>0.2ps, <6ns) collected with the commercial spectrometer. Relative intensity of the triplet peaks in >1ns region was adjusted to match intermediate time delay dataset (Figure 2a). To balance the signal coming from the PIA of singlet and triplet excitons, probe polarization was oriented approximately 60° relatively to the 0B crystal axis for the fastest (<10ps) and intermediate (>0.2ps, <6ns) setups measurements. A liquid nitrogen cryostat (OptistatDN-V, Oxford Instruments) was used for the temperature control.

# Section I. Matching the crystal-axis orientations with the habitus of rubrene single crystals

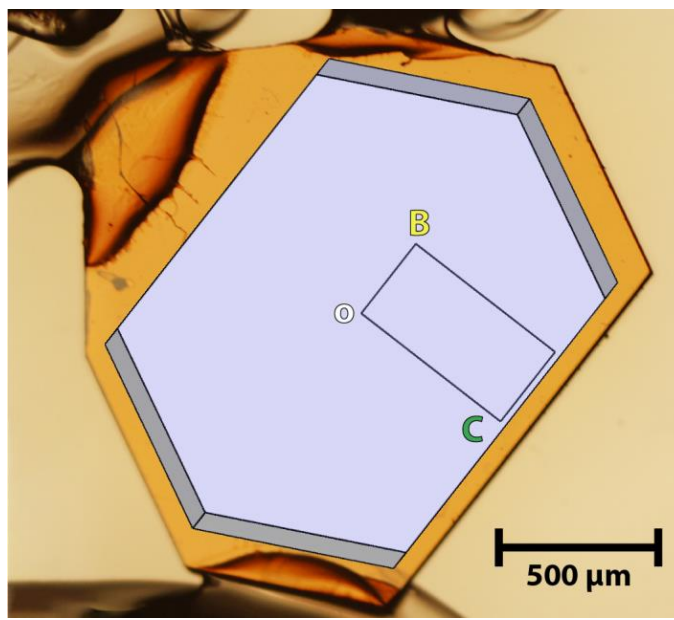

*Fig. S1. Assignment of the single crystal habitus to unit cell orientation. Background – a rubrene single crystal microscopy image. Foreground – habitus model and corresponding orientations of the unit cell axes calculated with the Morphology module of the Material Studio package[4] (Compass force field).[5]*

**Section II. Widths and positions of singlet and triplet spectral components as a function of temperature**

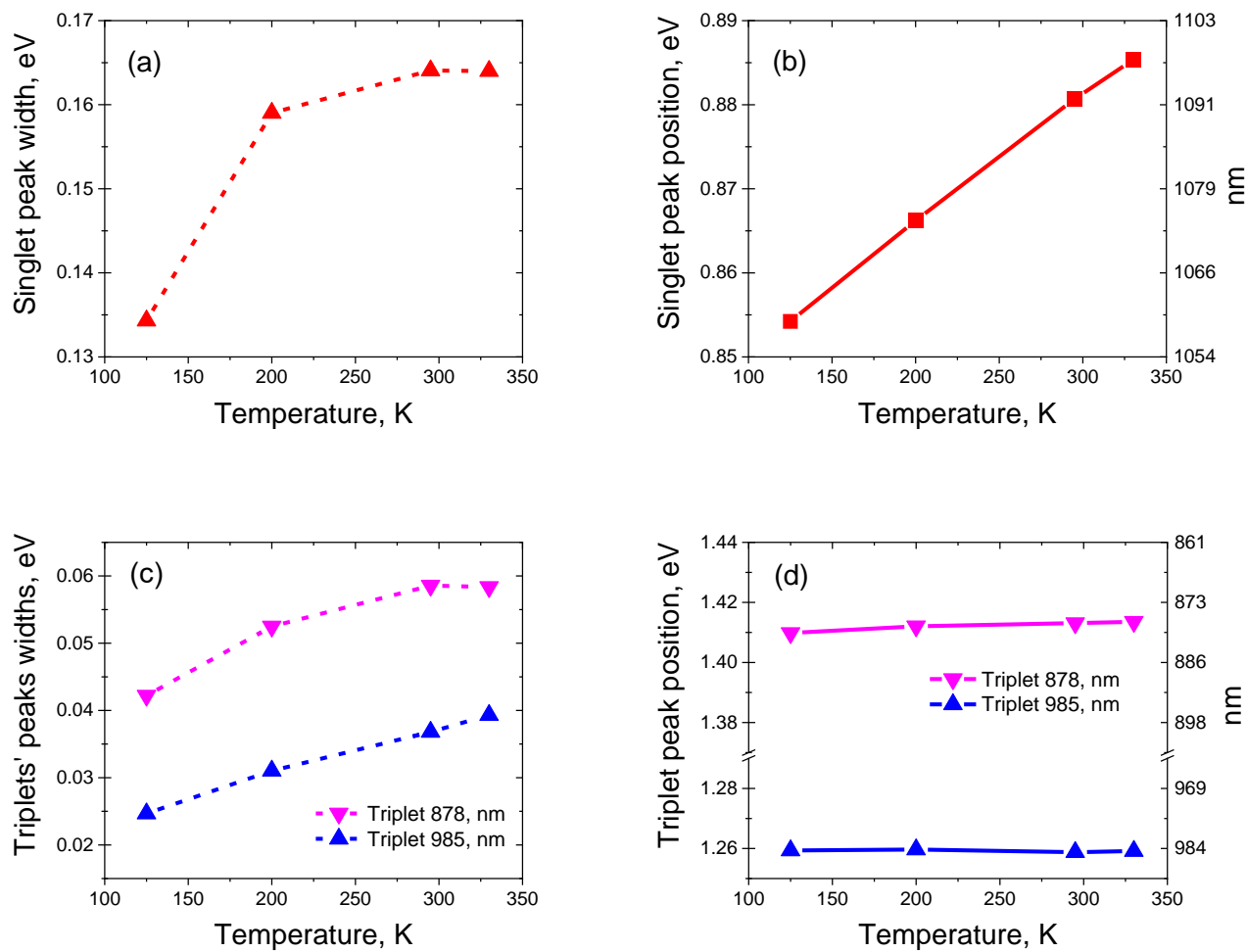

*Fig. S2. Spectral signatures of singlet and triplet components as a function of temperature. Peaks widths (a, c) and peaks positions (b, d).*

## Section III. Computational methodology

The ground state ($S_0$) geometry of the rubrene molecule was optimized at the density functional theory (DFT) level using the B3LYP functional and 6-31G(d,p) basis set. Additionally, harmonic vibrational frequencies were calculated on the energy-minimized structure to ensure that the obtained geometry does not contain any imaginary frequencies. Subsequent time-dependent DFT (TD-DFT) calculations were carried out to optimize the geometries of the first singlet ($S_1$) and triplet ($T_1$) excited states. The Tamm-Dancoff approximation (TDA)[6] scheme was taken into account during the TDDFT calculations to avoid any instabilities related to triplet states. To determine the electron-vibrational couplings for the $S_0 \rightarrow T$ and $S_1 \rightarrow T$ transitions, we computed the Huang-Rhys (HR) factors between the relevant states. The computed HR factors are summarized in Tables S2-S4. All DFT calculations were performed with the Gaussian 16 package.[7] (We note that we initially aimed to utilize the long-range corrected ωB97X-D functional with an optimally tuned range-separation parameter to calculate the HR factors; however, we encountered technical issues that prevented us from getting reliable results.)

The S $\rightarrow$ TT transition rate constant (k) was estimated within the Marcus-Levich-Jortner[8] model:

$$k = \frac{2\pi}{\hbar} t^2 \times \frac{1}{\sqrt{4\pi\lambda_c k_B T}} \sum_{n=0}^{\infty} \frac{e^{-S_{qm}} \times S_{qm}^n}{n!} e^{\frac{-(\Delta E + \lambda_c + n\hbar\omega_{qm})^2}{4\lambda_c k_B T}} \tag{S1}$$

where t denotes electronic coupling; $k_B$, the Boltzmann constant; $\hbar$, the reduced Plank constant; $\Delta E$, the energy difference between $S_0+S$ and $T+T$ states; $\lambda_c$, the classical reorganization energy with low frequency modes (i.e., <100 cm-1); $\hbar\omega_{qm}$, the vibration energy of the effective high-frequency quantum mode; and $S_{qm}$, the Huang-Rhys factor associated with the reorganization energy from high-frequency modes.

We estimated $\omega_{qm}$ and $S_{qm}$ as follows:

$$\omega_{qm} = \sqrt{\frac{(\sum \lambda_i \omega_i^2)_{S_0 \rightarrow T_1} + (\sum \lambda_i \omega_i^2)_{S_1 \rightarrow T_1}}{(\sum \lambda_i)_{S_0 \rightarrow T_1} + (\sum \lambda_i)_{S_1 \rightarrow T_1}}} \tag{S2}$$

$$\lambda_{qm} = \hbar\omega_{qm} S_{qm} \tag{S3}$$

where $\omega_i$ and $\lambda_i$ are the frequency and the related reorganization energy of the vibrational mode *i*, while $\lambda_{qm}$ denotes the reorganization energy due to high-frequency (quantum) vibrational modes. For the rate calculations, we considered the experimental temperature (T) range, from 125 K to 330 K. The S and T state energies were taken from reported experimental data as 2.23 eV and 1.14 eV, respectively, resulting in a $\Delta E$ value of 0.05 eV.[9] The low-frequency vibrational modes (see Tables S2-S4) with frequencies below 100 cm$^{-1}$ were treated as classical, while the high-frequency modes (>100 cm$^{-1}$) were treated quantum mechanically. This leads to $\omega_{qm}$=1368 cm$^{-1}$, $\lambda_c$=185 cm$^{-1}$, and $S_{qm}$=1.64. To take into account the medium effects and crystal lattice relaxations, we added an amount of 0.1 eV to $\lambda_c$.

In the 125 K – 330 K temperature range, the experimental rates of singlet fission (SF) show a variation from ~ 7×10$^9$ s$^{-1}$ to 1.35×10$^{11}$ s$^{-1}$, as illustrated in Figure 3f in the main text. In order to ensure that the computational SF rate derived in this study using the Marcus-Levich-Jortner equation (S1) aligns with the experimental values, it is crucial to select an appropriate electronic coupling in addition to the above-mentioned parameters. Therefore, we opted to use a small electronic coupling of either 60 cm$^{-1}$ (when 0.1 eV is not added to $\lambda_c$) or 95 cm$^{-1}$ (when 0.1 eV is added to $\lambda_c$), which ensures consistency of the computed rates with the experimental values.

Table S1 displays the SF rates calculated for different temperature ranges. By fitting the natural log (ln) of the SF rate vs $1/k_BT$, we obtain the activation energy ($E_a$) for $S_0S_1 \rightarrow T_1T_1$ to be 49.4 meV (Fig. S3). The inclusion of the environmental effects and crystal lattice relaxations to the classical reorganization energy ($\lambda_c$) increases $E_a$ to 52.2 meV.

*Table S1. SF rates at different temperatures as calculated using the Marcus-Levich-Jortner equation at the B3LYP/6-31G(d,p) level of theory. Rate-I corresponds to the case when 0.1 eV is not added to λc, while Rate-II corresponds to the case when 0.1 eV is added to λc.*

| T (K) | Rate-I (s⁻¹) | Rate-II (s⁻¹) |
|-------|--------------|----------------|
| 125 | $8.44 \times 10^9$ | $7.02 \times 10^9$ |
| 150 | $1.89 \times 10^{10}$ | $1.64 \times 10^{10}$ |
| 200 | $5.02 \times 10^{10}$ | $4.61 \times 10^{10}$ |
| 250 | $8.8 \times 10^{10}$ | $8.36 \times 10^{10}$ |
| 300 | $1.26 \times 10^{11}$ | $1.22 \times 10^{11}$ |
| 330 | $1.47 \times 10^{11}$ | $1.45 \times 10^{11}$ |

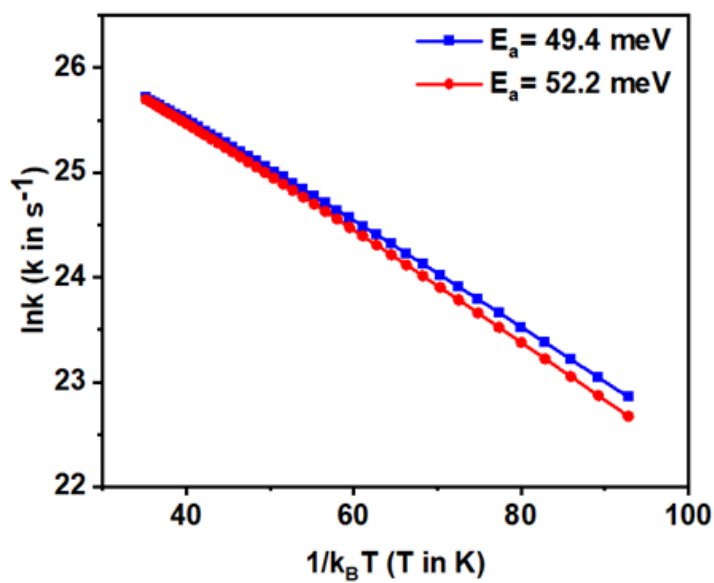

*Fig. S3. Rate constants (k) as a function of inverse temperature (T) between 125 K and 330 K. The blue line is generated by adding 0.1 eV to λc to incorporate the effects of the medium and crystal lattice relaxations, while the red line does not consider such effects.*

*Table S2. Relaxation energies and Huang-Rhys (HR) factors related to the $S_0 \rightarrow S_1$ transition as computed at the B3LYP/6-31G(d,p) level of theory.*

| $S_0 \rightarrow S_1$ | | | $S_1 \rightarrow S_0$ | | |
|---|---|---|---|---|---|
| Frequencies ($\omega$ in $cm^{-1}$) | Relaxation energy (in $cm^{-1}$) | (HR) factor | Frequencies ($\omega$ in $cm^{-1}$) | Relaxation energy (in $cm^{-1}$) | (HR) factor |
| 23 | 249.00 | 10.826 | 27 | 354.20 | 13.119 |
| 69 | 42.80 | 0.620 | 67 | 6.30 | 0.094 |
| 80 | 27.60 | 0.345 | 76 | 67.30 | 0.886 |
| 134 | 0.20 | 0.001 | 133 | 0.70 | 0.005 |
| 214 | 82.60 | 0.386 | 218 | 1.60 | 0.007 |
| 253 | 1.60 | 0.006 | 249 | 30.90 | 0.124 |
| 265 | 4.10 | 0.015 | 264 | 1.20 | 0.005 |
| 339 | 53.10 | 0.157 | 338 | 81.10 | 0.240 |
| 416 | 1.80 | 0.004 | 416 | 4.80 | 0.012 |
| 450 | 5.30 | 0.012 | 432 | 2.20 | 0.005 |
| 535 | 4.30 | 0.008 | 526 | 9.10 | 0.017 |
| 624 | 3.80 | 0.006 | 616 | 50.70 | 0.082 |
| 631 | 16.90 | 0.027 | 627 | 35.30 | 0.056 |
| 659 | 0.60 | 0.001 | 643 | 0.10 | 0.000 |
| 674 | 270.10 | 0.401 | 670 | 36.40 | 0.054 |
| 718 | 2.10 | 0.003 | 717 | 2.10 | 0.003 |
| 767 | 0.30 | 0.000 | 757 | 2.80 | 0.004 |
| 781 | 0.50 | 0.001 | 770 | 0.70 | 0.001 |
| 858 | 1.70 | 0.002 | 857 | 0.00 | 0.000 |
| 909 | 149.10 | 0.164 | 910 | 415.30 | 0.456 |
| 924 | 28.80 | 0.031 | 924 | 33.70 | 0.036 |
| 969 | 0.10 | 0.000 | 939 | 22.30 | 0.024 |
| 977 | 4.40 | 0.005 | 972 | 1.40 | 0.001 |
| 993 | 14.00 | 0.014 | 992 | 1.10 | 0.001 |
| 1002 | 175.60 | 0.175 | 1014 | 220.10 | 0.217 |
| 1017 | 363.60 | 0.358 | 1015 | 248.80 | 0.245 |
| 1057 | 173.00 | 0.164 | 1058 | 180.80 | 0.171 |
| 1067 | 12.40 | 0.012 | 1088 | 0.80 | 0.001 |
| 1104 | 1.20 | 0.001 | 1107 | 2.70 | 0.002 |
| 1186 | 3.70 | 0.003 | 1186 | 3.50 | 0.003 |
| 1201 | 9.90 | 0.008 | 1197 | 73.50 | 0.061 |
| 1208 | 115.00 | 0.095 | 1208 | 110.80 | 0.092 |
| 1232 | 72.30 | 0.059 | 1243 | 54.10 | 0.044 |
| 1319 | 38.20 | 0.029 | 1319 | 4.60 | 0.003 |
| 1347 | 234.40 | 0.174 | 1354 | 8.10 | 0.006 |
| 1353 | 14.20 | 0.010 | 1360 | 1.50 | 0.001 |
| 1357 | 27.10 | 0.020 | 1372 | 224.60 | 0.164 |
| 1477 | 53.00 | 0.036 | 1423 | 196.80 | 0.138 |
| 1482 | 8.70 | 0.006 | 1481 | 7.90 | 0.005 |
| 1529 | 14.50 | 0.009 | 1530 | 38.80 | 0.025 |
| 1542 | 10.50 | 0.007 | 1540 | 93.00 | 0.060 |
| 1592 | 236.80 | 0.149 | 1590 | 175.40 | 0.110 |
| 1629 | 3.20 | 0.002 | 1624 | 3.90 | 0.002 |
| 1656 | 5.80 | 0.004 | 1649 | 21.60 | 0.013 |
| 3174 | 14.60 | 0.005 | 3175 | 21.80 | 0.007 |
| 3182 | 0.40 | 0.000 | 3182 | 0.90 | 0.000 |
| 3192 | 66.50 | 0.021 | 3194 | 49.00 | 0.015 |
| 3197 | 0.00 | 0.000 | 3200 | 10.90 | 0.003 |
| 3199 | 14.50 | 0.005 | 3201 | 0.30 | 0.000 |
| 3207 | 269.60 | 0.084 | 3207 | 280.60 | 0.087 |
| 3240 | 1.40 | 0.000 | 3238 | 1.30 | 0.000 |
| Total relaxation energy | | | | | |
| | 0.361 eV | | | 0.397 eV | |

*Table S3. Relaxation energies and Huang-Rhys (HR) factors related to the $S_0 \rightarrow T_1$ transition as computed at the B3LYP/6-31G(d,p) level of theory.*

| $S_0 \rightarrow T_1$ | | | $T_1 \rightarrow S_0$ | | |
|---|---|---|---|---|---|
| Frequencies ($\omega$ in cm$^{-1}$) | Relaxation energy (in cm$^{-1}$) | (HR) $S_0 \rightarrow T_1$ | Frequencies ($\omega$ in cm$^{-1}$) | Relaxation energy (in cm$^{-1}$) | (HR) $T_1 \rightarrow S_0$ |
| 23 | 101.00 | 4.391 | 23 | 104.30 | 4.535 |
| 69 | 6.00 | 0.087 | 66 | 0.10 | 0.002 |
| 80 | 5.90 | 0.074 | 78 | 8.30 | 0.106 |
| 134 | 22.50 | 0.168 | 128 | 17.70 | 0.138 |
| 214 | 31.70 | 0.148 | 215 | 6.50 | 0.030 |
| 253 | 4.70 | 0.019 | 250 | 17.80 | 0.071 |
| 265 | 12.30 | 0.046 | 266 | 10.40 | 0.039 |
| 339 | 83.40 | 0.246 | 335 | 108.70 | 0.324 |
| 416 | 0.00 | 0.000 | 415 | 0.00 | 0.000 |
| 450 | 16.90 | 0.038 | 437 | 5.50 | 0.013 |
| 535 | 6.80 | 0.013 | 529 | 10.50 | 0.020 |
| 624 | 0.30 | 0.000 | 609 | 6.10 | 0.010 |
| 631 | 0.40 | 0.001 | 630 | 19.50 | 0.031 |
| 659 | 1.20 | 0.002 | 641 | 2.20 | 0.003 |
| 674 | 70.10 | 0.104 | 673 | 2.20 | 0.003 |
| 718 | 0.30 | 0.000 | 717 | 0.20 | 0.000 |
| 767 | 1.60 | 0.002 | 765 | 7.20 | 0.009 |
| 781 | 7.50 | 0.010 | 772 | 1.00 | 0.001 |
| 858 | 0.20 | 0.000 | 858 | 0.10 | 0.000 |
| 909 | 0.30 | 0.000 | 912 | 82.20 | 0.090 |
| 924 | 1.60 | 0.002 | 925 | 4.60 | 0.005 |
| 969 | 0.40 | 0.000 | 949 | 12.60 | 0.013 |
| 977 | 4.60 | 0.005 | 972 | 0.60 | 0.001 |
| 993 | 11.30 | 0.011 | 993 | 0.00 | 0.000 |
| 1002 | 138.80 | 0.139 | 1015 | 15.40 | 0.015 |
| 1017 | 28.10 | 0.028 | 1018 | 42.20 | 0.041 |
| 1057 | 14.10 | 0.013 | 1058 | 19.90 | 0.019 |
| 1067 | 2.20 | 0.002 | 1094 | 0.80 | 0.001 |
| 1104 | 0.10 | 0.000 | 1106 | 3.90 | 0.004 |
| 1186 | 0.40 | 0.000 | 1186 | 0.50 | 0.000 |
| 1201 | 54.40 | 0.045 | 1196 | 121.80 | 0.102 |
| 1208 | 11.60 | 0.010 | 1208 | 9.20 | 0.008 |
| 1232 | 7.30 | 0.006 | 1238 | 1.90 | 0.002 |
| 1319 | 68.80 | 0.052 | 1303 | 259.50 | 0.199 |
| 1347 | 488.90 | 0.363 | 1324 | 32.50 | 0.025 |
| 1353 | 94.60 | 0.070 | 1357 | 6.30 | 0.005 |
| 1357 | 91.40 | 0.067 | 1367 | 395.60 | 0.289 |
| 1477 | 22.20 | 0.015 | 1405 | 185.40 | 0.132 |
| 1482 | 0.30 | 0.000 | 1480 | 6.00 | 0.004 |
| 1529 | 103.70 | 0.068 | 1529 | 140.10 | 0.092 |
| 1542 | 52.50 | 0.034 | 1540 | 115.00 | 0.075 |
| 1592 | 402.20 | 0.253 | 1611 | 195.80 | 0.122 |
| 1629 | 0.50 | 0.000 | 1627 | 0.00 | 0.000 |
| 1656 | 0.40 | 0.000 | 1654 | 2.80 | 0.002 |
| 3174 | 1.30 | 0.000 | 3175 | 1.60 | 0.001 |
| 3182 | 0.00 | 0.000 | 3183 | 0.10 | 0.000 |
| 3192 | 6.60 | 0.002 | 3193 | 5.40 | 0.002 |
| 3197 | 0.20 | 0.000 | 3199 | 0.70 | 0.000 |
| 3199 | 1.80 | 0.001 | 3200 | 0.80 | 0.000 |
| 3207 | 27.70 | 0.009 | 3207 | 28.00 | 0.009 |
| 3240 | 0.10 | 0.000 | 3235 | 0.40 | 0.000 |
| Total relaxation energy | | | | | |
| | 0.249 eV | | | 0.251 | |

*Table S4. Relaxation energies and Huang-Rhys (HR) factors related to the $S_1{\rightarrow}T_1$ transition as computed at the B3LYP/6-31G(d,p) level of theory.*

| $S_1{\rightarrow}T_1$ | | | $T_1{\rightarrow}S_1$ | | |
|---|---|---|---|---|---|
| Frequencies ($\omega$ in cm$^{-1}$) | Relaxation energy (in cm$^{-1}$) | (HR) $S1{\rightarrow}T1$ | Frequencies ($\omega$ in cm$^{-1}$) | Relaxation energy ( in cm$^{-1}$) | (HR) $T1{\rightarrow}S1$ |
| 27 | 47.30 | 1.752 | 23 | 34.70 | 1.509 |
| 67 | 4.80 | 0.072 | 66 | 6.30 | 0.095 |
| 76 | 19.90 | 0.262 | 78 | 12.80 | 0.164 |
| 133 | 16.70 | 0.126 | 128 | 17.30 | 0.135 |
| 218 | 0.00 | 0.000 | 215 | 1.70 | 0.008 |
| 249 | 0.00 | 0.000 | 250 | 0.20 | 0.001 |
| 264 | 2.90 | 0.011 | 266 | 2.20 | 0.008 |
| 338 | 3.70 | 0.011 | 335 | 4.20 | 0.013 |
| 416 | 4.20 | 0.010 | 415 | 2.30 | 0.006 |
| 432 | 7.90 | 0.018 | 437 | 5.50 | 0.013 |
| 526 | 0.20 | 0.000 | 529 | 0.60 | 0.001 |
| 616 | 2.70 | 0.004 | 609 | 1.60 | 0.003 |
| 627 | 0.10 | 0.000 | 630 | 5.60 | 0.009 |
| 643 | 1.90 | 0.003 | 641 | 4.10 | 0.006 |
| 670 | 3.30 | 0.005 | 673 | 5.30 | 0.008 |
| 717 | 0.10 | 0.000 | 717 | 0.10 | 0.000 |
| 757 | 1.30 | 0.002 | 765 | 0.60 | 0.001 |
| 770 | 1.00 | 0.001 | 772 | 1.20 | 0.002 |
| 857 | 0.20 | 0.000 | 858 | 0.50 | 0.001 |
| 910 | 4.20 | 0.005 | 912 | 18.30 | 0.020 |
| 924 | 1.50 | 0.002 | 925 | 0.40 | 0.000 |
| 939 | 0.00 | 0.000 | 949 | 0.70 | 0.001 |
| 972 | 0.00 | 0.000 | 972 | 0.00 | 0.000 |
| 992 | 0.40 | 0.000 | 993 | 0.10 | 0.000 |
| 1014 | 47.60 | 0.047 | 1015 | 2.60 | 0.003 |
| 1015 | 1.70 | 0.002 | 1018 | 19.10 | 0.019 |
| 1058 | 6.00 | 0.006 | 1058 | 8.50 | 0.008 |
| 1088 | 0.00 | 0.000 | 1094 | 0.00 | 0.000 |
| 1107 | 0.80 | 0.001 | 1106 | 1.30 | 0.001 |
| 1186 | 0.00 | 0.000 | 1186 | 0.10 | 0.000 |
| 1197 | 3.80 | 0.003 | 1196 | 6.60 | 0.006 |
| 1208 | 5.70 | 0.005 | 1208 | 4.50 | 0.004 |
| 1243 | 6.80 | 0.005 | 1238 | 0.40 | 0.000 |
| 1319 | 10.70 | 0.008 | 1303 | 98.70 | 0.076 |
| 1354 | 34.50 | 0.025 | 1324 | 18.40 | 0.014 |
| 1360 | 46.00 | 0.034 | 1357 | 0.10 | 0.000 |
| 1372 | 45.20 | 0.033 | 1367 | 10.20 | 0.007 |
| 1423 | 0.20 | 0.000 | 1405 | 7.30 | 0.005 |
| 1481 | 0.30 | 0.000 | 1480 | 0.60 | 0.000 |
| 1530 | 18.00 | 0.012 | 1529 | 21.80 | 0.014 |
| 1540 | 35.40 | 0.023 | 1540 | 11.30 | 0.007 |
| 1590 | 16.40 | 0.010 | 1611 | 12.50 | 0.008 |
| 1624 | 0.10 | 0.000 | 1627 | 0.00 | 0.000 |
| 1649 | 2.20 | 0.001 | 1654 | 0.00 | 0.000 |
| 3175 | 1.00 | 0.000 | 3175 | 0.60 | 0.000 |
| 3182 | 0.00 | 0.000 | 3183 | 0.00 | 0.000 |
| 3194 | 1.80 | 0.001 | 3193 | 2.30 | 0.001 |
| 3200 | 0.40 | 0.000 | 3199 | 0.20 | 0.000 |
| 3201 | 0.00 | 0.000 | 3200 | 0.20 | 0.000 |
| 3207 | 10.20 | 0.003 | 3207 | 9.90 | 0.003 |
| 3238 | 0.00 | 0.000 | 3235 | 0.20 | 0.000 |
| Total relaxation energy | | | | | |
| | 0.052 eV | | | 0.045 eV | |

*Vibronic model*

In order to simulate concomitantly the absorption spectrum of the rubrene molecule and to account for the **S:TT** coupling, we used a 3-state model comprising the ground state and the $S_1$ and $^1TT$ states:

$$H_{vib} = \begin{pmatrix} E_G + H_{vib}^{(G)} & -(\vec{d}\vec{E}) & 0 \\ -(\vec{d}\vec{E}) & E_S + H_{vib}^{(S)} & t \\ 0 & t & E_{TT} + H_{vib}^{(TT)} \end{pmatrix} \tag{S4}$$

$$H_{vib}^{(G)} = \sum_i \frac{\hbar\omega_i}{2} q_i^2 \tag{S5}$$

$$H_{vib}^{(S)} = \sum_i \frac{\hbar\omega_i}{2} q_i^2 + \sum_i \sqrt{2}\ \hbar\omega_i g_i^{(S)} + \lambda_s \tag{S6}$$

$$H_{vib}^{(TT)} = \sum_i \frac{\hbar\omega_i}{2} q_i^2 + \sum_i \sqrt{2}\ \hbar\omega_i g_i^{(TT)} + \lambda_{TT} \tag{S7}$$

$$\lambda = \sum_i \hbar\omega_i\ g_i^2 \tag{S8}$$

Here, $E_G$, $E_S$ and $E_{TT}$ are, respectively, the energies of the ground state and the S and TT states; **t** is the coupling between the S and TT states; $\omega_i$ and $g_i$ are, respectively, the frequency and the coupling constant of the vibrational mode q$_i$. $H_i = -(\vec{d}\vec{E})$ describes the coupling of the $S_0$ to S electronic excitation with the applied electromagnetic field (light) in the dipole approximation. This interaction is treated here perturbatively.

The dynamic solutions (Eq. S4) of the vibronic Hamiltonian were obtained numerically using a procedure described in our previous work.[10] The vibronic eigenfunctions are given as:

$$\psi_\alpha = \chi_\alpha^{(G)}(q)\psi^{(G)} + \chi_\alpha^{(S)}(q)\psi^{(S)} + \chi_\alpha^{(TT)}(q)\psi^{(TT)} \tag{S9}$$

Here, $\psi^{(M)}$ denotes the electronic wavefunction of state M and $\chi_\alpha^{(M)}(q)$ is the related vibrational function where q stands for the normal coordinates of all involved vibrational modes. The dynamic solutions of the vibronic Hamiltonian can be obtained by expending the functions $\chi_\alpha^{(M)}(q)$ in terms of harmonic-oscillator eigen-functions. By using a large but finite number of these functions in the expansion, one can obtain the vibronic solutions with any desirable accuracy.

If only the transitions from the ground vibronic state are considered, the absorption spectrum is proportional to $P_{0\alpha}^{(S)} = \left| \chi_0^{(G)}(q) \, \chi_\alpha^{(G)}(q) \right|^2$. The weight of the **TT** state to vibronic state α is given by $P_\alpha^{(TT)} = \left| \chi_\alpha^{(TT)}(q) \, \chi_\alpha^{(TT)}(q) \right|^2$. As a consequence, the product $P_{0\alpha} * P_\alpha^{(TT)}$ gives the probability of generating a TT state by means of photoexcitation. The probability, $(P^{(S)} * P^{(TT)})(E)$, of generating a TT state with energy E, is obtained by broadening the vibronic transition energies (using Gaussian functions) and summing up $P_{0\alpha} * P_\alpha^{(TT)}$ over all transitions. Finally, the probability of generating a TT state per absorbed photon with energy E, (referred here as the TT photogeneration population) was computed by dividing $(P^{(S)} * P^{(TT)})(E)$ by $P^{(S)}(E)$.

In the vibronic calculations, we used two effective vibrational modes, one high-frequency mode (ω₁) and one low-frequency vibrational mode (ω₂). The parameters in the $H_{vib}^{(S)}$ term (Eq. S6) were chosen such that the calculations reproduce the experimental absorption spectrum of isolated rubrene molecules. The coupling to a low energy (ω₂=200 cm⁻¹) vibrational mode was considered only in the $H_{vib}^{(TT)}$ term (Eq. S7). The vibrational constants entering $H_{vib}^{(TT)}$ were chosen to reproduce the reorganization energy related to **S** →**TT** transition.

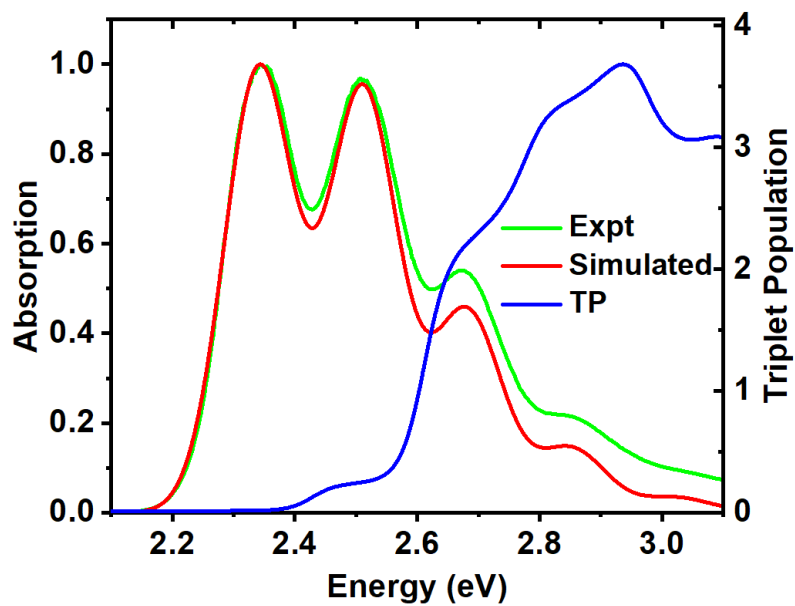

*Fig. S4. S1 absorption of rubrene in deuterated chloroform solution (green curve), its simulation (red curve) and the relative weight (population) of the **TT** state normalised per absorbed photon at a given energy (blue curve). The vibronic Hamiltonian was solved by using the following microscopic parameters: t = 7.4 meV , $\omega_1$=1370 cm$^{-1}$ , $\omega_2$=200 cm$^{-1}$, $g_1^{(S)}$= -0.975, $g_2^{(S)}$ = 0, $g_1^{(TT)}$ = 0.1, and $g_2^{(TT)}$ = 4.*

# Supplementary References: